\documentclass[aps,prl,twocolumn, superscriptaddress]{revtex4-1}
\usepackage[T1]{fontenc}
\usepackage{amsmath}
\usepackage{amssymb}
\usepackage{nicefrac}
\usepackage{siunitx}
\usepackage{graphicx}
\usepackage[tracking]{microtype}

\newcommand{\avr}[1]{\langle#1\rangle}
\newcommand{\LT}{\mathsf{LT}}

\SetTracking[outer kerning={0,0}, spacing={-30*,,}]{encoding={*}, shape=sc}{30}

\begin{document}
\title{Asymptotically Exact Solution of the Fredrickson-Andersen Model}

\author{Koray \"Onder}
\email{Koray.Oender@dlr.de}
\affiliation{Institut f\"{u}r Theoretische Physik, Universit\"at zu K\"oln, 50937 K\"{o}ln, Germany}
\affiliation{Institut f\"{u}r Materialphysik im Weltraum, Deutsches Zentrum f\"{u}r Luft- und Raumfahrt (DLR), 51170 K\"{o}ln, Germany}

\author{Matthias Sperl}
\affiliation{Institut f\"{u}r Materialphysik im Weltraum, Deutsches Zentrum
  f\"{u}r Luft- und Raumfahrt (DLR), 51170 K\"{o}ln, Germany}
\affiliation{Institut f\"{u}r Theoretische Physik, Universit\"at zu K\"oln, 50937 K\"{o}ln, Germany}

\author{W. Till Kranz}
\affiliation{Institut f\"{u}r Theoretische Physik, Universit\"at zu K\"oln, 50937 K\"{o}ln, Germany}

\date{\today}
\begin{abstract}
  The Fredrickson-Andersen (\textsc{fa}) model---a kinetically constrained
  lattice model---displays an ergodic to non-ergodic transition with a slow
  two-step relaxation of dynamical correlation functions close to the
  transition point. We derive an asymptotically exact solution for the
  dynamical occupation correlation function of the \textsc{fa} model on the
  Bethe lattice by identifying an exact expression for its memory kernel. The
  exact solution fulfills a scaling relation between critical exponents and
  allows to calculate the exponents explicitly. In addition, we propose an
  approximate dynamics that describes numerical data away from the critical
  point over many decades in time.
\end{abstract}


\maketitle

Slow relaxation is not restricted to molecular fluids dominated by pairwise
interactions. On the contrary, systems abound where the effective dynamics is
facilitated by the number of neighbors in a favorable state exceeding a
threshold. Dynamic facilitation applies to opinion dynamics
\cite{ramos+shao15}, voter models \cite{castellano+munoz09,jkedrzejewski17},
and infection spreading \cite{chae+yook15} but has also been used to
understand the low temperature phase of magnetic alloys
\cite{pollak+riess75,Chalupa1979}, granular compaction \cite{brey+prados99},
rigidity percolation \cite{moukarzel+duxbury97}, and the jamming transition
\cite{toninelli+biroli06,schwarz+liu06}. Most prominently it lies at the heart
of the dynamic facilitation picture
\cite{glarum60,fredrickson+andersen84,garrahan+chandler02,evans02,ritort+sollich03,chandler+garrahan10}
of the glass transition
\cite{angell+ngai00,berthier+biroli11,hunter+weeks12,biroli+garrahan13}. The
k-core decomposition of graphs \cite{seidman83,dorogovtsev+goltsev06} yields a
statistical description of, \textit{e.g.}, social groups \cite{seidman83} and
the brain \cite{turova12}. K-core decomposition can be framed as a dynamic
facilitation problem \cite{baxter+dorogovstev15} yielding, \textit{e.g.},
insight into the resilience of social network data sets against
de-anonymization \cite{yartseva+grossglauser}.

A paradigmatic example of a kinetically constraint model
\cite{ritort+sollich03,garrahan+sollich11} implementing dynamic
facilitation is the Fredrickson-Andersen (\textsc{fa}) model which is
defined on a lattice with sites $i=1,\ldots,N$ decorated with
occupation numbers $n_i\in\{0,1\}$. The Hamiltonian
$\mathcal H = \mu\sum_in_i$ is trivial and $\mu > 0$ favors the empty
lattice. A site $i$ may, however, only change its state if it has at
least $f$ empty nearest neighbors
\cite{fredrickson+andersen84,cancrini+martinelli09}. Bootstrap
percolation \cite{Chalupa1979,adler91,dorogovtsev+goltsev08,saberi15}
is concerned with the ground state of the \textsc{fa} model that is
kinetically reachable from an initial condition with an occupation
probability $p$. For $p=1$, clearly the occupation probability in the
ground state $q=1$, whereas for $p\approx0$ an empty ground state,
$q=0$, can be reached almost surely. The question arises, if there is
a nontrivial concentration, $p_c$, for the emergence of an infinite
occupied cluster in the ground state, $q>0$. For $f=1$ and for
arbitrary $f$ on hypercubic lattices $\mathbb Z^d$ it has been shown
that $p_c=1$ \cite{vanenter87,cancrini+martinelli09}. Bootstrap
percolation on the Bethe lattice and on random graphs, however,
feature a transition at a finite $p_c < 1$
\cite{Chalupa1979,balogh+pittel07,janson+luczak12}.

At finite temperatures $T > 0$ \footnote{Temperature is measured in
  units of $|\mu|/k_B$} we equip the \textsc{fa} model with transition
rates that satisfy detailed balance. Without constraints, the
Hamiltonian $\mathcal H$ would entail an equilibrium mean occupation
$\avr{n_i} = 1/(1 + e^{1/T})$. This still holds under the constrained
dynamics as long as $\avr{n_i} < p_c$, however, for $\avr{n_i} > p_c$,
the dynamics is restricted to the sites that are not permanently
constrained by the frozen percolating cluster
\cite{cancrini+martinelli08}. For $\avr{n_i}\nearrow p_c$, numerical
simulations of the \textsc{fa} model
\cite{Sellitto2005,arenzon+sellitto12,Sellitto2015,decandia+fierro16}
show a two-step relaxation of time-correlation functions, $\phi(t)$,
with a fast relaxation to a plateau value, $\phi(t)\simeq q > 0$,
followed by a second relaxation, $\phi(t)\to0$, on a time scale that
diverges towards $p_c$. A two-step relaxation with a divergent
relaxation time is one of the experimental fingerprints of the glass
transition \cite{angell+ngai00} and motivated the \textsc{fa} model as
an effective description of the glass transition. Close to the
plateau, $|\phi(t) - q| \ll 1$, the relaxation is generically well
described by power laws \cite{goetze+sjoegren92},
\begin{equation}
  \label{eq:9}
  \phi(t) - q \propto\left\{
    \begin{aligned}
      t^{-a} & \text{ for } \phi(t) > q,\\
      -t^b &  \text{ for } \phi(t) < q.
    \end{aligned}\right.
\end{equation}

\begin{figure}[t]
  \centering
  \includegraphics{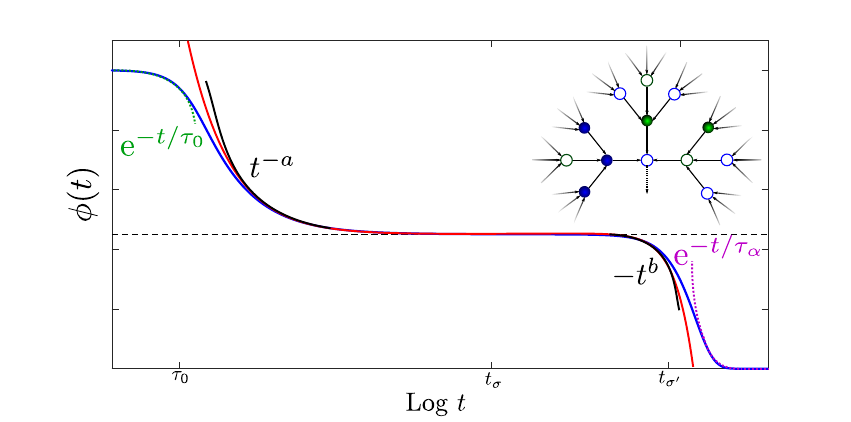}
  \caption{Generic occupation correlation $\phi(t)$ (blue) of the
    oriented \textsc{fa} model (\textsc{ofa}) close to the critical
    point, $\sigma\ll1$, compared to the asymptotic scaling function
    $G_{\sigma}(t) \simeq \phi(t) - q_c$ [Eq.~(\ref{eq:15}), red] and
    the critical laws $t^{-a}$ and $-t^b$ [Eq.~(\ref{eq:9}), black].
    The plateau $q_c$ (horizontal dashed line) is crossed, on a
    divergent timescale $t_{\sigma} \sim \sigma^{-\delta}$ and
    $G_{\sigma}(t)$ provides a faithful description in a divergent
    time window $\tau_0\ll t\ll t'_{\sigma} \sim \sigma^{-\gamma}$
    beyond the microscopic relaxation time $\tau_0$, followed by
    exponential relaxation on a timescale
    $\tau_{\alpha}\sim t'_{\sigma}$.  \textit{Inset:} Section of the
    \textsc{ofa} for coordination $k=3$, and facilitation $f=2$.
    Arrows denote the orientation, filled (open) vertices denote
    occupied (empty) sites, and blue vertices are frozen while green
    may change. See Table~\ref{tab:fa32} for values of the critical
    exponents.}
  \label{fig:schematic}
\end{figure}

A complementary description of the glass transition, independent of
the dynamic facilitation picture, is provided by mode-coupling theory
(\textsc{mct}) \cite{cummins99,Goetze2008,janssen18} which starts
from the formally exact equation of motion
\begin{equation}
  \label{eq:8}
  \tau_0\dot \phi(t) + \phi(t) + \int_0^t\mathrm{d}t'm(t-t')\dot \phi(t') = 0,
\end{equation}
where the dot denotes the time derivative and $\tau_0^{-1}$ is the short
time relaxation rate. The eponymous mode-coupling approximation
(\textsc{mca}) expresses the unknown memory kernel $m(t)$ by a
polynomial in $\phi(t)$. Standard \textsc{mct} predicts a scaling
relation,
\begin{equation}
  \label{eq:10}
  \frac{\Gamma^2(1 - a)}{\Gamma(1 - 2a)} 
  = \frac{\Gamma^2(1 + b)}{\Gamma(1 + 2b)},
\end{equation}
between the exponents in Eq.~(\ref{eq:9}) involving the Euler
Gamma-function. \textsc{mca}s have been attempted for the \textsc{fa} model
\cite{eisinger+jaeckle93,pitts+young00,pitts+andersen01,einax+schulz01,schulz+trimper02}
starting with \citet{fredrickson+andersen84} but were of limited success. In
particular, \textsc{mct} for the \textsc{fa} model has a tendency to predict
spurious transitions \cite{ritort+sollich03}. Also other approaches did not
capture the slow relaxation \cite{fennell+gleeson14}.

Recent numerical evidence, however has shown that despite these reservations,
the scaling relation~(\ref{eq:10}) seems to be verified in the \textsc{fa}
model on the Bethe lattice modeled as a random regular graph (\textsc{rrg})
\cite{Sellitto2005,sellitto+demartino10,arenzon+sellitto12,Sellitto2015,decandia+fierro16}. Proof
for this surprising discovery is highly desired
\cite{cancrini+martinelli09,rizzo18} but missing so far.

\begin{table}[t]
  \caption{Critical exponents $a,b,\delta=1/2a,\gamma=(1/2a) + (1/2b)$
    and exponent parameter $\lambda$ characterizing the
    slow relaxation of the \textsc{fa} model for coordination $k=3$
    and facilitation $f=2$ (see Fig.~\ref{fig:schematic}). Comparing
    the  exact results derived here to fits from simulation data.}
  \centering
  \begin{tabular*}{\columnwidth}{c@{\extracolsep{\fill}}SSS}
    \toprule\vspace{-6pt}\\
 & {Theory} & \multicolumn{2}{c}{Simulation}\\
 & & {\citet{Sellitto2015}} & {\citet{decandia+fierro16}}\\\hline
    $\lambda$ & {$\nicefrac{839}{864}$\footnote{$\lambda=0.971\ldots$}} & 0.815 & 0.79\\\
    $a$ & 0.121\ldots & 0.27 & 0.29\\
    $b$ & 0.147\ldots & 0.45 & 0.50\\
    $\delta$ & 4.13\ldots & 1.85 & 1.72\\
    $\gamma$ & 7.53\ldots & 2.96 & 2.72\\
    \botrule
  \end{tabular*}
  \label{tab:fa32}
\end{table}

In this letter we derive an asymptotically exact solution of
Eq.~(\ref{eq:8}) for the \textsc{fa} model on the Bethe lattice. We
show that Eq.~(\ref{eq:9}) constitutes the lowest order in a series
expansion of this solution and that the scaling relation~(\ref{eq:10})
holds exactly. Encouraged by these results we propose an approximate,
regularized memory kernel valid for all times. Comparing with
numerical data far away from the critical point, we are able to
describe the two step relaxation of the \textsc{fa} model over many
decades in time. 

\textit{Model.---}We consider the oriented \textsc{fa} model
(\textsc{ofa}) with facilitation parameter $f\geq2$ on the Bethe
lattice. To be precise, we define the Bethe lattice \cite{Mezard2001}
as the infinite $k$-ary rooted tree \cite{Martinelli2013},
$2\leq f < k$. In line with Sellitto's numerical work
\cite{Sellitto2005,arenzon+sellitto12,Sellitto2015} we assume
Metropolis dynamics with transition rates
$w(n_i \to 1-n_i) = \exp[(n_i-1)/T]C_f(\mathcal K_i)$.  Here
$\mathcal K_i$ denotes the set of children of site $i$ and
$C_f:\mathcal K_i\mapsto\{0,1\}$ implements the kinetic constraint
\footnote{$C_f(\mathcal K_i) = \Theta\big(k - f + \nicefrac12 -
  \sum_{j\in\mathcal K_i}n_i\big)$, with the Heaviside
  step-function $\Theta(x)$.}  (cf.\ Fig.~\ref{fig:schematic}).

For simplicity we aim to describe the relaxation to equilibrium from a
well defined initial condition. Assume the initial $n_i(0)$ are drawn
from a Bernoulli distribution with $p = 1/(1 + e^{-1/T}) > 1/2$
\footnote{Think of this distribution as the unconstrained model's
  equilibrium for $\mu < 0$}. To assure ergodicity, we limit our
discussion to $p < p_c$.

\textit{Percolation Transition.---}Recall that the probability, $q$,
that a site is occupied in the ground state can be given implicitly as
\cite{Chalupa1979}
\begin{equation}
  \label{eq:12}
  q = pQ(q) := p\sum_{i=0}^{f-1}
  \begin{pmatrix}
    k\\i
  \end{pmatrix}
  q^{k-i}(1-q)^i.
\end{equation}
Note that as $p > q$, the largest real solution of Eq.~(\ref{eq:12}) is
physically relevant. Trivially, $q=0$ is always a solution of
Eq.~(\ref{eq:12}). The critical probability $p_c$ locates a bifurcation to
additional solutions. Generically, Eq.~(\ref{eq:12}) displays a fold
bifurcation (Arnold's type A$_2$ \cite{Arnold1992}) with a finite $q_c \equiv
q(p_c) > 0$ and close to the transition $q(p>p_c) - q_c \sim \sqrt{p - p_c}$.

\textit{Equation of Motion.---}We wish to describe the single site
occupation correlation function
\begin{equation}
  \label{eq:11}
  \phi(t) = \avr{ n_s(0) n_s(t)},
\end{equation}
where $n_s$ denotes the occupation number of an arbitrary but fixed
site $s$ and the average $\avr\cdot$ is taken with respect to the
initial distribution. Note that $\phi(t)$ is normalized such that
$\phi(0) = p$ and $\phi(t\to\infty) = q$.

For $p < p_c$, the \textsc{ofa} is a Markov process obeying detailed
balance. Hence standard techniques allow to give the time evolution of the
distribution function in terms of an effective Hamiltonian $H$
\cite{risken96}. Applying a Mori projector, $n_s\rangle\avr{n_s^2}^{-1}\langle
n_s$, and rewriting the memory kernel in terms of its irreducible counterpart,
$m(t)$, yields Eq.~(\ref{eq:8}) \cite{Kawasaki1995}. The rate $\tau_0^{-1}$
can be calculated explicitly \footnote{$\tau_0^{-1} = p -
  p(1-p)^k\sum_{i=0}^{f-1}\binom ki[p/(1-p)]^i$}.  The memory kernel, however,
is only known formally.

\textit{Critical Dynamics.---}It is instructive to rewrite
Eq.~(\ref{eq:8}) in the Laplace domain, $\hat\phi(z) = \LT[\phi](z)$
\footnote{$\LT[\phi](z) :=
  i\int_0^{\infty}\phi(t)e^{izt}\mathrm{d}t$}. For $z\ll \tau_0^{-1}$
one finds
\begin{equation}
  \label{eq:13}
  \frac{z\hat\phi(z)}{p + z\hat\phi(z)} = z\hat m(z).
\end{equation}
In particular $q/(p - q) = m(t\to\infty)$. Comparing this with
Eq.~(\ref{eq:12}) we arrive at our central result: Asymptotically the
memory kernel of the \textsc{ofa} on the Bethe lattice is given
exactly as
\begin{equation}
  \label{eq:14}
  m(t\to\infty) \equiv m(p,q) = pQ(q)/(p - q).
\end{equation}

Sufficiently close to the critcial point,
$\sigma := (p_c - p)/p_c \ll 1$, we expect a growing window in time,
centered around a diverging time scale $t_{\sigma}$ where
$\phi(t) = q_c + G_{\sigma}(t/t_{\sigma})$ such that $G_{\sigma}(t)$
is small, $|G_{\sigma}(t/t_{\sigma})|\ll1$, and slowly varying,
$|\zeta\hat G_{\sigma}(\zeta)|\ll1$, where $\zeta := zt_{\sigma}$. To
this end we expand Eq.~(\ref{eq:13}) around $q_c$ to lowest order in
$G_{\sigma}$ \cite{goetze84,Goetze2008},
\begin{equation}
  \label{eq:15}
  \lambda\LT[G_{\sigma}^2](\zeta) + \zeta\hat G_{\sigma}^2(\zeta) 
  = -\sigma/\zeta p_c,
\end{equation}
where
$\lambda := 1 + (p_c - q_c)^3\partial_q^2\Delta m(p_c, q_c)/2p_c < 1$
\footnote{$\Delta m(p,q) := m(p,q) - q/(p - q)$}.  As $\lambda$ can be
calculated exactly for the \textsc{ofa}, the same holds for
$G_{\sigma}(t)$. Eq.~(\ref{eq:15}) can be solved by standard numerical
techniques (cf.\ Fig.~\ref{fig:schematic}) but more information can be
gained analytically.

At the critical point, $\sigma\equiv0$, Eq.~(\ref{eq:15}) is solved by
$G_0(t) \sim t^x$, provided $\lambda = \Gamma^2(1+x)/\Gamma(1 + 2x)$
\cite{goetze84}. Asymptotically, $G_0(t\to\infty) \sim t^{-a}$, the
smallest negative $x\equiv-a<0$ will dominate. Away from the critical
point, for finite $\sigma$, $G_{\sigma}(t) = \sqrt{\sigma}g(t)$
acquires a square-root dependence on $\sigma$. Eq.~(\ref{eq:15}) still
admits power law solutions, $g(t) \sim t^x$, iff the left hand side
dominates over the right hand side. For the approach to the plateau,
$t\to t_{\sigma}$, $g(t/t_{\sigma}) \sim (t/t_{\sigma})^{-a}$, as long
as $(t/t_{\sigma})^{2a}\ll1$. Matching
$G_{\sigma\searrow0}(t/t_{\sigma}) = G_0(t)$ for $t\to\infty$, yields,
$t_{\sigma} \sim \sigma^{-\delta}$, where $\delta := 1/2a$
\cite{goetze84}. For $\lambda>1/2$ and times $t > t_{\sigma}$, the
decay away from the plateau is governed by the smallest positive
$x\equiv b$, $g(t/t_{\sigma}) \sim -(t/t_{\sigma})^b$, as soon as
$g^2(t/t_{\sigma})\gg1$, i.e., dependent on $b<1/2$ ($b>1/2$) for
times $(t/t_{\sigma})^{2b} \gg 1$ ($t/t_{\sigma} \gg 1$). For long
times the validity of this law is limited by the slowly varying
condition,
$|\zeta\hat G_{\sigma}(\zeta)| \sim \sqrt{\sigma}\zeta^{-b} \ll 1$,
i.e., for times $t \ll t'_{\sigma} \sim \sigma^{-\gamma}$, where
$\gamma := (1/2a) + (1/2b)$ \cite{goetze84}.

The above constitutes a precise statement of Eq.~(\ref{eq:9}) for the
\textsc{ofa} and proves that the scaling relation~(\ref{eq:10}) holds
exactly. We summarize the (numerically) exact results we obtain for
the simplest model, $k=3$ and $f=2$ in Tab.~\ref{tab:fa32}.

\textit{Asymptotic Relaxation.---} The asymptotic relaxation to zero,
$\phi(t\to\infty)\to0$, on times $t/t'_{\sigma}\gg1$ is governed by a
scaling function, $\phi(t\to\infty) = \tilde\phi(t/t'_{\sigma})$
\cite{goetze84}. For the \textsc{ofa} it has been shown that
$\phi(t) \propto \exp(-t/\tau_{\alpha})$ where
$\tau_{\alpha} \sim \sigma^{-\gamma'}$ and the exponent
$\gamma' \ge 2$ could only be bounded from below
\cite{cancrini+martinelli15}. In terms of the scaling function we find
$\tau_{\alpha}\sim t'_{\sigma}$ and in particular the preceding
analysis determines the exponent $\gamma'\equiv\gamma$ compatible with
the bound.

\textit{Persistence Function.---}The persistence function $\psi(t)$
yields the fraction of sites that have not changed their state since
$t=0$. Its asymptotic value, $\psi_{\infty} = \psi(t\to\infty)$, does
not only include the persistently occupied sites but also a fraction
of the empty sites that are permanently frozen,
$\psi_{\infty}(q) = pQ(p) + (1-p)Q(p) = q/p$. Close to the plateau,
$\psi_{\infty}^c = q_c/p_c$, we can expand
\begin{equation}
  \label{eq:17}
  \psi(t\sim t_{\sigma}) 
  \simeq \psi_{\infty}^c + \partial_q\psi_{\infty}(q_c)[\phi(t) - q_c],
\end{equation}
i.e., $\psi(t) \simeq \psi_{\infty}^c + \sqrt{\sigma}g(t)/p_c$.  In
particular, the persistence function is governed by the same critical
exponents and master function $g(t)$ that apply to $\phi(t)$.

\textit{The FA Model on Random Regular Graphs.---}It is known that
bootstrap percolation on the oriented and unoriented Bethe lattice of
coordination $k+1$ \cite{Martinelli2013} as well as on random
$(k+1)$-regular graphs \cite{balogh+pittel07} have the same critical
concentration $p_c$. Not much is known regarding the dynamic
equivalence. Here we conjecture that due to the bifurcation dominating
close to the critical point, the fact that the critical point does not
change translates to dynamic equivalence close to $q_c$. For the
unoriented \textsc{fa} model the expression for the persistence
function, $\tilde\psi_{\infty}(q)$, is slightly more involved
\cite{Sellitto2005}. Nevertheless,
$\partial_q\tilde\psi_{\infty}(q_c)$ is finite and therefore
Eq.~(\ref{eq:17}) applies and $\tilde\psi(t)$ is still governed by the
critical exponents and scaling function of the \textsc{ofa}.

Simulations
\cite{Sellitto2005,Sellitto2015,decandia+fierro16,arenzon+sellitto12,sellitto13}
of the \textsc{fa} model are conveniently being performed on
\textsc{rrg}s with a finite number of sites $N$. \textsc{rrg}s do not,
however, admit an orientation. The effective system size is given by
the size of the largest embedded tree $L = O(\log_kN)$
\cite{math/0610858} which grows with $N$ but is still small even for
$N\simeq 2^{24}$.  Therefore the existing numerical data is relatively
far from the critical point. In addition the \textsc{fa} model is
known to display strong finite size effects \cite{aizenman+lebowitz88}
which, so far, have not been analyzed in detail for \textsc{rrg}s. As
a consequence, the empirical critical exponents (Table~\ref{tab:fa32})
are effective exponents and deviate from the analytical predictions.
In the following we propose a memory kernel that allows us to solve
Eq.~(\ref{eq:8}) for all times and for appreciable distances $\sigma$
from the critical point relevant to the numerical data.

\begin{figure}[t]
  \begin{center}
    \includegraphics{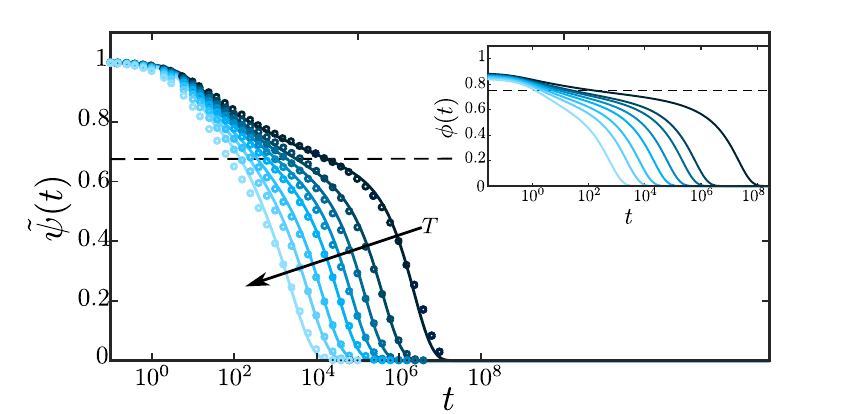}
  \end{center}
  \caption{\label{fig:exact} Persistence function $\tilde\psi(t)$ of
    the unoriented \textsc{fa} model as a function of time for
    coordination $k=3$ and facilitation $f=2$ at a number of
    temperatures $T = 0.49$, 0.5, 0.51, 0.52, 0.54, 0.56, 0.58, 0.6,
    above the critical temperature $T_c = 1/\ln 8 = 0.480\ldots$ The
    dashed line indicates the asymptotic plateau value
    $\tilde\psi_{\infty}^c = \frac{2757}{4096} = 0.673\ldots$
    \cite{Sellitto2005}. Symbols are simulations from
    Ref.~\cite{Sellitto2015} while lines are the theoretical
    prediction, Eq.~(\ref{eq:18}). \textit{Inset:} Corresponding
    theoretical occupation correlation functions $\phi(t)$,
    Eqs.~(\ref{eq:8}, \ref{eq:16}). The dashed line indicates the
    plateau value $q_c = 3/4$.}
\end{figure}

\textit{Approximate Memory Kernel.---}To close Eq.~(\ref{eq:8}) we
propose to approximate the memory kernel by Eq.~(\ref{eq:14}) for all
times as $m(t)\approx \tilde m[\phi(t)] \equiv m(p, \phi(t))$.
Unfortunately this is not viable as
$\tilde m[\phi(t)\to p] \sim [p - \phi(t)]^{-1}$ diverges for small
times, $\phi(t\to0)\to p$. In order to regularize the memory kernel we
assume the divergent term,
$[p - \phi(t)]^{-1} = p^{-1}\sum_i^M [\phi(t)/p]^i, M\to\infty$, to be
a resummation of many-site interactions. On a finite lattice, the
order of interactions should be finite, $M < \infty$. Therefore we
propose a regularized approximate memory kernel whose time-dependence
is completely determined by $\phi(t)$,
\begin{equation}
  \label{eq:16}
  m(t) \approx m[\phi(t)] = Q(\phi(t))\sum_{i=0}^M[\phi(t)/p]^i.
\end{equation}
With this Eq.~(\ref{eq:8}) can be numerically solved for $\phi(t)$ by
standard techniques \cite{fuchs+goetze91}.

\textit{Discussion.---}Considering the occupation correlation function
$\phi(t)$, Eq.~(\ref{eq:11}), of the \textsc{ofa} on the Bethe
lattice, we have identified an explicit expression, Eq.~(\ref{eq:14}),
for the long time limit of its memory kernel. Expanding around the
bifurcation at $p_c$ that signals the ergodic to non-ergodic
transition of the \textsc{ofa}, we find that close to the transition,
$\sigma\to0$, the time evolution of $\phi(t)$ around its plateau value
$q_c$ is asymptotically exactly given,
$\phi(t) = q_c + \sqrt{\sigma}g(t)$, in terms of a one-parameter
scaling function $g(t)\equiv g_{\lambda}(t)$, Eq.~(\ref{eq:15}). The
exponent parameter $\lambda \equiv \lambda(k, f)$ is known explicitly
in terms of the lattice coordination $k$ and the facilitation
parameter $f$. The properties of Eq.~(\ref{eq:15}), finally, imply
Eq.~(\ref{eq:9}) together with the scaling relation~(\ref{eq:10}), for
$\lambda > 1/2$, times
$\tau_0\ll t \ll t'_{\sigma} \sim \sigma^{-\gamma}$, and not too close
to the plateau, $|\phi(t) - q| \gg \sigma$.

The scaling function $G_{\sigma}(t)$, however, goes beyond
Eq.~(\ref{eq:9}) as it provides a faithful description of $\phi(t)$
for $\tau_0 \ll t \ll t'_{\sigma}$, $\lambda > 0$, bounded only by the
requirements $|G_{\sigma}(t)|, |\zeta\hat G_{\sigma}(\zeta)|\ll1$
(cf.\ Fig.~\ref{fig:schematic}). On the fast end this could be
complemented by ever more sophisticated short-time expansions. On the
long-time end, $\phi(t\to\infty) \sim e^{-t/\tau_{\alpha}}$, with a
relaxation time $\tau_{\alpha}\sim\sigma^{-\gamma}$.

Considering the asymptotic dynamics only, we did not gain information
about processes on intermediate time scales. Could we have missed an
additional process that will always mask the bifurcation scenario? The
answer is no: Any unidentified process must occur on a time scale,
$\tau_u$, that remains finite as $\sigma\to0$. Otherwise it would
contribute to Eq.~(\ref{eq:12}). Therefore we can always find a
$\sigma_0 > 0$ such that for $\sigma < \sigma_0$,
$t_{\sigma} \sim \sigma^{-\delta} \gg \tau_u$ and we have a time
window which is dominated by the bifurcation.

Given that close to the critical point the persistence function,
$\psi(t)$, is governed by the same scaling function $G_{\sigma}(t)$,
Eq.~(\ref{eq:17}) provides an asymptotically exact description of the
persistence function in a divergent time window before the asymptotic
exponential relaxation. The form of Eq.~(\ref{eq:9}) and
the scaling relation~(\ref{eq:10}) equally apply to $\psi(t)$ with the
qualifications given above. Thereby we confirm the empirical
observation of \citet{Sellitto2015} and \citet{decandia+fierro16}.

To close Eq.~(\ref{eq:8}), we proposed a memory functional,
Eq.~(\ref{eq:16}), regularized by a finite length scale $M$ we
conjecture to be related to the system size. Formally,
Eq.~(\ref{eq:16}) looks like a \textsc{mca} but let us stress that it
was not derived by considering a (physically motivated) coupling of
modes, but ultimately from the bifurcation equation~(\ref{eq:12}) of
the underlying bootstrap percolation.

To determine the persistence function of the unoriented \textsc{fa}
model for all times, we use the knowledge gained so far and
interpolate
\begin{multline}
  \label{eq:18}
  \tilde\psi(t) = e^{-t/\tau_{\psi}} + \left\{
    \tilde\psi_{\infty}^c + \partial_q\tilde\psi_{\infty}(q_c)[\phi(t) - q_c]
  \right\}\times\\
  \times\big(1 - e^{-t/\tau_{\psi}}\big)e^{-t/\tau_{\alpha}},
\end{multline}
where $\tau_{\psi}^{-1}$ is the short time relaxation rate of
$\tilde\psi(t)$. We determine $\phi(t)$ by solving Eq.~(\ref{eq:8})
with the regularized memory kernel [Eq.~(\ref{eq:16})] and treat $M$
as a fit parameter \footnote{Fitted $M=6$--23 increase with
  $\sigma\to0$ as expected.}. Fig.~\ref{fig:exact} shows excellent
agreement between Eq.~(\ref{eq:18}) and the numerical data for all
temperatures and over many decades in time. 

The success of this approach, derived for the \textsc{ofa} on the
Bethe lattice, in describing simulations of the unoriented \textsc{fa}
model on \textsc{rrg}s provides reasons to assume that the similarity
between the oriented and unoriented \textsc{fa} model extends beyond a
common critical point $p_c$ to a universal dynamics close to $p_c$. It
is, however, obvious, that simulations much closer to the critical
point are needed to challenge the conjectures put forward here and to
confirm the critical exponents.

While for sake of brevity we have only presented explicit results for
the simplest case, $k=3, f=2$, our approach holds for more general
coordinations $k > 3$, and facilitation parameters $f\ge 2$ provided
$k > f$. The consequences of a tunable $\lambda(k, f)$ will be
discussed elsewhere, but let us note that for some combinations
$(k, f)$, $\lambda < \lambda(3, 2)$. As a result the exponents $a,b$
increase which may be favorable for simulations.

The signature of a fold bifurcation, $q - q_c \sim \sqrt{|\sigma|}$,
with a finite critical $q_c > 0$ is observed as a hybrid phase
transition in a variety of models
\cite{silbert+liu05,cai+chen15,cho+lee16}. A similar analysis to the
one introduced here could lead to new insights in those systems as
well. Reconciling \textsc{mct} and replica methods led to many new
insights provided by the random first order theory (\textsc{rfot})
\cite{wolynes+lubchenko12}. A deeper analysis of the overlap between
\textsc{mct} and dynamic facilitation theory that has been started
here and \textsc{rfot} and dynamic facilitation
\cite{foini+krzakala12} is likely to provide additional understanding
of the glass transition.

In summary we have provided an asymptotically exact description of the
slow relaxation of the oriented Fredrickson-Andersen model on the
Bethe lattice close to its critical point, valid over a divergent
window in time. We believe our method can be applied to other time
correlation functions of the \textsc{fa} and related kinetically
constraint models and can provide new insights into the phenomena
which can be mapped onto these models.

\begin{acknowledgments}
  We are indebted to Wolfgang G\"otze for posing the initial questions
  that led to this work. We thank Mauro Sellitto for sharing his
  simulation data and acknowledge additional insight from discussions
  with Thomas Franosch and Thomas Voigtmann. Partial funding was
  provided by the \textsc{dfg} through KR\,4867/2-1 (W.T.K.) and by
  the BMWi through 50\,WM\,1651 (K.\"O.).
\end{acknowledgments}

%

\end{document}